\documentclass[12pt]{article}
\begin{document}

\input{psfig}

\def \ms {{\overline{\mbox{MS}}}}
\newcommand{\z}{&&\hspace*{-0.7cm}}
\newcommand{\zz}{&&\hspace*{-0.3cm}}
\newcommand{\bea}{\begin{eqnarray}}
\newcommand{\eea}{\end{eqnarray}}
\newcommand{\prepr}[1] {\begin{flushright} {\bf #1} \end{flushright} 
\vskip 1.5cm}

\begin{flushright} {
%4rs version\\
\bf US-FT/23-00 \\ 
%December 20th, 2000
} \end{flushright}

%Small $x$ 
\begin{center}
{\bfseries
$Q^2$ EVOLUTION OF PARTON DISTRIBUTIONS AT SMALL $X$.\\
A STUDY OF HIGHER-TWIST EFFECTS}

\vskip 5mm

A.V. Kotikov$^{1 \dag}$ and  G. Parente$^{2 \dag} $ 

\vskip 5mm

{\small
(1) {\it
Particle Physics Laboratory,
Joint Institute for Nuclear Research,~
141980 Dubna, Russia
}
\\
(2) {\it
%Dep.
Departamento 
de F\'\i sica de Part\'\i culas, 
%Univ.
Universidade 
de Santiago de Compostela,
\\
15706 Santiago de Compostela, Spain
}
\\
$\dag$ {\it
E-mail: kotikov@sunse.jinr.ru, gonzalo@fpaxp1.usc.es
}}
\end{center}

\vskip 5mm

\begin{center}
\begin{minipage}{150mm}
\centerline{\bf Abstract}

We investigate  the  $Q^2$ evolution of parton distributions
at small $x$ values,
%present an analytical parametrization of the QCD
%description of the  behaviour
%of parton distribution functions in the leading twist approximation
%of the Wilson operator product expansion,
%recently 
obtained in the case
of flat initial conditions.
The contributions of twist-two and (renormalon-type) higher-twist 
operators of the Wilson operator product expansion are taken into account.
The results are in excellent agreement with deep inelastic scattering
experimental data from HERA.
\end{minipage}
\end{center}

\vskip 10mm

\section{Introduction}

The measurements of the deep-inelastic scattering
%(DIS)
structure function
%(SF)
$F_2$ in HERA
%\cite{H1,ZEUS,ZEUSB}
\cite{H1,H1n}
have permitted the access to
a very interesting kinematical range for testing the theoretical
ideas on the behavior of quarks and gluons carrying a very low fraction
of momentum of the proton, the so-called small $x$ region.
In this limit one expects that
non-perturbative effects may give essential contributions. However, the
reasonable agreement between HERA data and the 
%NLO 
next-to-leading order (NLO)
approximation of
perturbative
QCD that has been observed for $Q^2 > 1 $GeV$^2$ (see the recent review
in \cite{CoDeRo}) indicates that
%and, thus,
perturbative QCD could describe the
evolution of structure functions up to very low $Q^2$ values,
traditionally explained by soft processes.
It is of fundamental importance to find out the kinematical region where
the well-established perturbative QCD formalism
can be safely applied at small $x$.

The standard program to study the small $x$ behavior of
quarks and gluons
is carried out by comparison of data
with the numerical solution of the
%Dokshitzer-Gribov-Lipatov-Altarelli-Parisi (DGLAP)
DGLAP
equations 
%\cite{DGLAP1, DGLAP}
%\footnote{ At small $x$ there is another approach
%based on the
%%Balitsky-Fadin-Kuraev-Lipatov (BFKL) equation \cite{BFKL},
%BFKL equation, whose
%application is out of the scope of this work.} 
by fitting the parameters of the
$x$ profile of partons at some initial $Q_0^2$ and
the QCD energy scale $\Lambda$ (see, for example, \cite{MRS,KKPS}).
%\cite{fits}-\cite{GRV}.
However, if one is interested in analyzing exclusively the
small $x$ region ($x \leq 0.01$), 
there is the alternative of doing a simpler analysis
by using some of the existing analytical solutions of DGLAP 
in the small $x$ limit (see \cite{CoDeRo} for review).
%\cite{BF1}-\cite{Munich2}.
This was done so in Ref. \cite{BF1}-\cite{Q2evo3}
where it was pointed out that the HERA small $x$ data can be
interpreted in 
terms of the so called doubled asymptotic scaling phenomenon
related to the asymptotic 
behavior of the DGLAP evolution 
discovered  in \cite{Rujula}
many years ago. 

Here we illustrate results obtained recently in
\cite{Q2evo}
and demonstrate some (preliminary) results of \cite{Q2evo3}, where
the contributions of higher-twist operators 
(i.e. twist-four ones and twist-six ones)
of
the Wilson operator product expansion are taken into account.
The importance of the contributions of higher-twist operators
at small-$x$ has been done in many studies (see \cite{Bartels}).

We would like to note that
the results of \cite{Q2evo}
are the extension to the NLO QCD approximation of previous 
leading order (LO)
%LO 
studies \cite{Rujula,BF1}.
The main ingredients are:

{\bf 1.} Both, the gluon and quark singlet densities are
presented in terms of two components ($'+'$ and $'-'$)
which are obtained from the analytical $Q^2$
dependent expressions of the corresponding ($'+'$ and $'-'$)
parton distributions moments.

{\bf 2.} The $'-'$ component is constant
at small $x$, whereas the 
$'+'$ component grows at $Q^2 \geq Q^2_0$ as 
%$$\sim \exp{\left(2\sqrt{\left[
%a_+\ln \left(
%\frac{a_s(Q^2_0)}{a_s(Q^2)} \right) -
%\left( b_+ +  a_+ \frac{\beta_1}{\beta_0} \right)
%\Bigl( a_s(Q^2_0) - a_s(Q^2) \Bigr) \right] 
%\ln \left( \frac{1}{x}  \right)} \right)},
%$$
$\sim \exp{(\sigma_{NLO})}$, where
\bea
\sigma_{NLO} = 2\sqrt{(\hat d_{+}s+\hat D_{+}p)lnx},
%\label{NLO}
\nonumber
\eea
%where 
and the LO term $\hat d_+ = -12/\beta_0$ and the NLO one 
$\hat D_{\pm}=\hat d_{\pm\pm}+\hat d_{\pm}\beta_1/\beta_0$ with
$\hat d_{\pm\pm} = 412f/(27\beta_0)$. 
Here the coupling constant
$a_s=\alpha_s/(4\pi)$, 
$s=ln(\alpha (Q^2_0)/\alpha (Q^2))$ and
$p=\alpha (Q^2_0) - \alpha (Q^2)$,
$\beta_0$ and $\beta_1$ are the first two 
coefficients of QCD 
$\beta$-function and $f$ is the number of active flavors.

\section{Basic formulae
%Approach
}

Our purpose
%of this article
is to show the small $x$ asymptotic
form of parton distributions
%(PD)
in the framework of the DGLAP equation starting at some $Q^2_0$ with
the flat function:
 \begin{eqnarray}
f^{\tau2}_a (Q^2_0) ~=~
A_a ~~~~(\mbox{ hereafter } a=q,g), \label{1}
 \end{eqnarray}
where $f^{\tau2}_a$ are the leading-twist parts of
parton (quark and gluon)
distributions multiplied by $x$
and $A_a$ are unknown parameters that have to be determined from data.
Through this work at small $x$ we neglect
the non-singlet quark component.

We would like to note that new HERA data \cite{H1n} show a rise
of $F_2$ structure function at low $Q^2$ values ($Q^2 < 1 $GeV$^2$)
when $x \to 0$ (see Fig.2, for example). The rise can be explained
in a natural way by incorporation  of higher-twist terms in our
analysis (see the part 2.2).

We shortly compile below the main results found in \cite{Q2evo,Q2evo3} 
at the LO
approximation (the leading-twist
%corresponding 
results at the NLO approximation
may be found in \cite{Q2evo}). 
%This will allow us to incorporate
%correctly the backward evolution, i.e. the possibility to go to $Q^2 < Q_0^2$,
%which was not considered in our previous work \cite{Q2evo}. 
%It will also permit to fix the notation which is used in the part 2.2
%%Section 3 
%to present, for the first time, the twist-4 contributions
%at low $x$.
%
%%Finally we present the 
The full small $x$ asymptotic results
for parton distributions
 and $F_2$ structure function at LO of perturbation theory is:
 \begin{eqnarray}
F_2(x,Q^2)&=& e \cdot f_q(x,Q^2)
\label{r10} \\ 
%& & \nonumber \\
 f_a(x,Q^2) &=& f^{+}_a(x,Q^2) + 
f^{-}_a(x,Q^2) \; , 
\label{r11}
\end{eqnarray}
where the $'+'$ and $'-'$ components $f^{\pm}_a(x,Q^2)$
%in the renormalon case 
are given by the sum
 \begin{eqnarray}
 f^{\pm}_a(x,Q^2) ~=~ f^{\tau2,\pm}_a(x,Q^2) + 
f^{h\tau,\pm}_a(x,Q^2) \;  
\label{r12}
\end{eqnarray}
of
the leading-twist parts $f^{\tau2,\pm}_a(x,Q^2)$ 
%(see Eqs. (\ref{8.0}) and (\ref{8.00})) 
and the 
higher-twist parts $f^{h\tau,\pm}_a(x,Q^2)$, 
%(see
%%$f^{ren,+}_q$,$f^{ren,+}_g$, $f^{ren,-}_q$ and 
%%$f^{ren,-}_g$ are given by 
%Eqs. (\ref{r5}) and (\ref{r9})), 
respectively.

%%%%%%%%%%%% 2.1 %%%%%%

\subsection{The contribution of twist-two operators} \indent

The small $x$ asymptotic results for PD, $f^{\tau2,\pm}_a$ 
%($a=q,g$) 
%and $F^{\tau2}_2$ structure function at LO of
%perturbation theory and at twist-two ($\tau2$)
%in the operator product expansion have been found in Ref. \cite{Q2evo}:
%
%\begin{eqnarray}
%F^{\tau2}_2(x,Q^2)&=& e \; f^{\tau2}_q(x,Q^2)
%\label{8} \\
%\nonumber \\
%f^{\tau2}_a(x,Q^2) &=& f^{\tau2,+}_a(x,Q^2) + 
%f^{\tau2,-}_a(x,Q^2) \;
%\label{8.2}
%\end{eqnarray}
%where
%$e= (\sum^f_1e^2_i)/f$ is the average charge square for $f$ active quarks.
%
%After Mellin inversion of the explicit moment solution to DGLAP
%equations, the $'+'$ and $'-'$ PD components are given by:
\begin{eqnarray}
f^{\tau2,+}_g(x,Q^2)&=& \biggl(A_g + \frac{4}{9} A_q \biggl)
\tilde I_0(\sigma) \; e^{-\overline d_{+}(1) s} ~+~O(\rho) 
~~\;\; ,\label{8.0} \\
f^{\tau2,+}_q(x,Q^2)&=& \frac{f}{9}\biggl(A_g + \frac{4}{9} A_q \biggl) 
\rho \; \tilde I_1(\sigma) \;
e^{-\overline d_{+}(1) s} ~+~O(\rho) \; , \label{8.01} 
%\nonumber
\\
%\nonumber \\ 
f^{\tau2,-}_g(x,Q^2)&=& - \frac{4}{9} A_q e^{- d_{-}(1) s} 
~+~O(x) ,
%~\mbox { and } 
\label{8.00} \\
f^{\tau2,-}_q(x,Q^2)&=&  A_q e^{- d_{-}(1) s} ~+~O(x) \; ,\label{8.02}
% \nonumber
 \end{eqnarray}
where
$\overline d_{+}(1) = 1+20f/(27\beta_0)$ and
$          d_{-}(1) = 16f/(27\beta_0)$
are the regular parts of $d_{+}$ and $d_{-}$
anomalous dimensions, respectively, in the limit $n\to1$ 
\footnote{From now on, for a quantity $k(n)$ we use the notation
$\hat k(n)$ for the singular part when $n\to1$ and
$\overline k(n)$ for the corresponding regular part. }. 
%
%We define the variable
% \begin{eqnarray}
%s=ln\left(\frac{a_s(Q^2_0)}{a_s(Q^2)}\right)
%\label{2.4}
% \end{eqnarray}
%
The functions $\tilde I_{\nu}$ ($\nu=0,1$) 
%in Eqs. (\ref{8.0,8.01}) 
are related to the modified Bessel
function $I_{\nu}$
and to the Bessel function $J_{\nu}$ by:
\begin{eqnarray}
\tilde I_{\nu}(\sigma) ~=~
%\[ 
\left\{
\begin{array}{rl}
I_{\nu}(\sigma), & \mbox{ if } s \geq 0 \\
J_{\nu}(\sigma), & \mbox{ if } s  <   0
\end{array} \right.  
%\]
.
\label{4}
\end{eqnarray}
The 
variables $\sigma$ and $\rho$ are
%argument $\sigma$ is 
given by
%\footnote{Hereafter we use the variables 
%$\sigma$ and $\rho$, introduced in Refs. \cite{BF1}
%for the case $Q^2 \geq Q_0^2$. In our article, they are generalized
%to arbitrary values of $Q^2$ and beyond
%the LO approximation (see Eq.(\ref{NLO}).}
\begin{eqnarray}
\sigma =2\sqrt{|\hat d_{+} s ln(x)|} \; , ~~~
\rho = \sqrt{\frac{|\hat d_{+} s|}{ln(1/x)}}
= \frac{\sigma}{2ln(1/x)}
\label{slo}
\end{eqnarray}
%where
%\begin{eqnarray}
%\hat d_{gg} = -4 C_A/\beta_0 
%\label{3.1d}
%\end{eqnarray}
%($C_A=N$ for $SU(N)$) is the singular part when $n \to 1$
%of $d_{gg}=\gamma^{(0)}_{gg}(n)/(2\beta_0)$ 
%being
%$\gamma^{(0)}_{gg}(n)$ 
%%and $\beta_0$ 
%the LO
%coefficient of the gluon-gluon AD.
%% and the QCD $\beta $-function respectively
%
%The prescription for the backward evolution given by Eq. (\ref{4})
%is the result, in the more general case, of the following representation
%of the series which appear in the
%inverse Mellin transformation of the exact solution for PD moments.
%(see for example Eq. (6) in Ref. \cite{twist2}),
%\begin{eqnarray}
%\sum^{\infty}_{k=0} \frac{t^k}{k!\Gamma(k+\nu +1)} ~=~
%{|t|}^{-\nu/2} \tilde I_{\nu}(2\sqrt{|t|}) \equiv
%%\[ 
%{|t|}^{-\nu/2}
%\left\{
%\begin{array}{rl}
%I_{\nu}(2\sqrt{|t|}), & \mbox{ if }t \geq 0 \\
%J_{\nu}(2\sqrt{|t|}), & \mbox{ if }t  <   0
%\end{array} \right.  
%%\]
%.
% \label{4.1}
% \end{eqnarray}
%
%Finally, in Eq. (\ref{8.0})
%\begin{eqnarray}
%\rho = \sqrt{\frac{|\hat d_{gg} s|}{ln(1/z)}}
%= \frac{\sigma}{2ln(1/z)}~,~~~ \label{rolo} 
%\end{eqnarray}
%%\overline \gamma^{(0)}_{gg}(1) = 
%%22 +\frac{4}{3}f ~~~ \mbox{ and } ~~~
%%\overline d_{gg}(1) = 1+\frac{4f}{3\beta_0}$$
%%with $f$ as the number of active quarks.

\subsection{The higher-twist contributions} \indent

%We consider renormalon model predictions for higher twist  operators.

Using the results in \cite{Q2evo3} (which are based on calculations
\cite{SMaMaS,method}), we show the effect
of higher-twist corrections in the renormalon case
(see recent review of renormalon models in \cite{Beneke}).
We present the results below 
%by using twist-two ones in 
%Eqs.(\ref{8.0})-(\ref{8.02}) and
making the following subtitutions
in the corresponding twist-two results presented in
Eqs.(\ref{8.0})-(\ref{8.02}):

%in the PD $f^{\tau2,+}_a(z,Q^2)$ ($a=q,g$) given by Eq.(\ref{8.0})
%(we start with consideration of LO evolution).
%
%In the gluon case 
%The modification  
$f^{\tau2,+}_g(x,Q^2)$ (see Eq.(\ref{8.0})) $\to f^{h\tau,+}_g(x,Q^2)$
by
%has the following form:
\bea 
A_a \tilde I_0(\sigma) &\to & A_a \cdot  
\frac{16f}{15\beta_0^2}  \Biggl\{ 
%& &
\frac{\Lambda^2_{1,a}}{Q^2} \left(
\frac{2}{\rho} \tilde I_1(\sigma)
+ \left[K_{ga}(f)
%\frac{101}{60}- \frac{8f}{81} 
-\ln \left(
\frac{\Lambda^2_{1,a}}{Q^2} \right)\right] 
\tilde I_0(\sigma)  \right)  \nonumber \\
&-&\frac{8}{7} \frac{\Lambda^4_{2,a}}{Q^4} \left(
\frac{2}{\rho} \tilde I_1(\sigma)
+ \left[K_{ga}(f)-
\frac{11}{112}
%- \frac{8f}{81} 
-\ln \left(
\frac{\Lambda^2_{2,a}}{Q^2} \right)\right] 
\tilde I_0(\sigma)  \right) \Biggl\}, 
%\Biggr] 
\label{r1}
\eea
where
$\Lambda^2_{1,a}$ and $\Lambda^4_{2,a}$ are magnitudes of twist-four and 
twist-six corrections and
\bea 
K_{gg}(f) ~=~ \frac{101}{60}- \frac{8f}{81},~~~
K_{gq}(f) ~=~ \frac{121}{60}- \frac{7f}{81};
%\nonumber \\
%K^{(2)}_{gg}(f) &=& \frac{2663}{1680}- \frac{8f}{81}, ~~ 
%K^{(2)}_{gq}(f) ~=~ \frac{3223}{1680}- \frac{7f}{81}
\nonumber
\eea

%In the quark case 
%The modification 
$f^{\tau2,+}_q(x,Q^2)$ (see Eq.(\ref{8.01})) $ \to f^{h\tau,+}_q(x,Q^2)$
by
%has the form:
\bea 
%\frac{4f}{81} 
A_a \rho \tilde I_1(\sigma) 
&\to & A_a \cdot
\frac{128f}{45\beta_0^2} \Biggl\{ 
%\nonumber \\& &
\frac{\Lambda^2_{1,a}}{Q^2} \left(
\frac{2}{\rho} \tilde I_1(\sigma)
+ \left[K_{qa}(f)
%\frac{11}{60}- \frac{2f}{27} 
-\ln \left(
\frac{\Lambda^2_{1,a}}{Q^2} \right)\right] 
\tilde I_0(\sigma)  \right)  \nonumber \\
&-&\frac{8}{7} \frac{\Lambda^4_{2,a}}{Q^4} \left(
\frac{2}{\rho} \tilde I_1(\sigma)
+ \left[K_{qa}(f)
-\frac{11}{112}
%- \frac{2f}{27} 
-\ln \left(
\frac{\Lambda^2_{2,a}}{Q^2} \right)\right] 
\tilde I_0(\sigma)  \right) \Biggl\}, 
%\Biggr] 
\label{r3}
\eea
where
\bea 
K_{qq}(f) ~=~ \frac{11}{60}- \frac{2f}{27},~~~
K_{qg}(f) ~=~ -\frac{3}{20}- \frac{7f}{81};
%\nonumber \\
%K^{(2)}_{qq}(f) &=& \frac{143}{1680}- \frac{2f}{81}, ~~ 
%K^{(2)}_{gq}(f) ~=~ -\frac{139}{1120}- \frac{7f}{81}
\nonumber
\eea

%The modification 
$f^{\tau2,-}_g(x,Q^2)$ (see Eq.(\ref{8.00}))
 $ \to f^{h\tau,-}_g(x,Q^2)$
by
%has the following form:
\bea 
%-\frac{4}{9}
A_q &\to & A_q \cdot
\frac{16f}{15\beta_0^2} \Biggl\{ 
%\nonumber \\ & &
\frac{\Lambda^2_{1,q}}{Q^2} \left(
\ln \left(\frac{Q^2}{x^2\Lambda^2_{1,q}} \right)
- \frac{33}{40}- \frac{7f}{81} \right)  
%\nonumber \\&-&
~-~\frac{8}{7} 
\frac{\Lambda^4_{2,q}}{Q^4} 
%\nonumber \\&\cdot&  
 \Biggl(
\ln \left(\frac{Q^2}{x^2\Lambda^2_{2,q}} \right)
\nonumber \\&+ & \frac{143}{1680}
- \frac{7f}{81} 
%-\ln \left(
%\frac{\Lambda^2_{2,q}}{Q^2} \right)
\Biggr) \Biggl\} 
%\Biggr] 
~-~
%\nonumber \\ &-& 
A_g \cdot
\frac{128f^2}{1215\beta_0^2}  \left\{
\frac{\Lambda^2_{1,g}}{Q^2} - \frac{8}{7}
 \frac{\Lambda^4_{2,g}}{Q^4} \right\};
\label{r6}
\eea

%In the quark case 
%The modification 
%of $f^{\tau2,-}_q(z,Q^2)$ 
$f^{\tau2,-}_q(x,Q^2)$ (see Eq.(\ref{8.02})) $ \to f^{h\tau,-}_q(x,Q^2)$
by
%becames:
\bea 
\z
A_q \to
%&\to & 
A_q \cdot
\frac{128f}{45\beta_0^2} \Biggl\{ 
%\nonumber \\ & &
\frac{\Lambda^2_{1,q}}{Q^2} \Biggl[
\biggl(\ln \left(\frac{Q^2}{x\Lambda^2_{1,q}} \right)
+ \frac{5}{2}
\biggl)
\ln \left(\frac{1}{x} \right)
-\left( \frac{359}{120}+ \frac{4f}{81}\right) 
\ln \left(\frac{Q^2}{x^2\Lambda^2_{1,q}} \right)
\nonumber \\
\zz
%&+&
~ +  \frac{1871}{600}+ \frac{309f}{1215}
+ \frac{24f^2}{(81)^2} 
%\left(\frac{1}{\delta_{q}}- \frac{139}{120}- \frac{4f}{81}\right)
%\ln \left(\frac{\Lambda^2_{1,q}}{Q^2} \right)
\Biggr]  
%\nonumber \\&-&
~ -
\frac{8}{7} \frac{\Lambda^4_{2,q}}{Q^4} \Biggl[
\biggl(\ln \left(\frac{Q^2}{x\Lambda^2_{2,q}} \right)
+ \frac{52}{21}
\biggl)
\ln \left(\frac{1}{x} \right)
\nonumber \\
\zz
%&-&
- 
\left( \frac{10237}{3360}+ \frac{4f}{81}\right) 
\ln \left(\frac{Q^2}{x^2\Lambda^2_{2,q}} \right) 
+\frac{33301}{11200}+ \frac{3377f}{19440} + \frac{24f^2}{(81)^2} 
%\nonumber \\&-&
%\left(
%\ln \left(\frac{Q^2}{x^2\Lambda^2_{2,q}} \right)
%
%\frac{1}{\delta_{q}}
%- \frac{3707}{3360}- \frac{4f}{81}\right)
%\ln \left(\frac{\Lambda^2_{2,q}}{Q^2} \right)
\Biggr] \Biggl\} 
%\Biggr] 
%\nonumber 
\label{r7}\\
\zz
%&-& 
-A_g \cdot
\frac{128f^2}{405\beta_0^2} \Biggl\{
\frac{\Lambda^2_{1,g}}{Q^2} 
%\nonumber \\&\cdot&
\left(
\ln \left(\frac{Q^2}{x^2\Lambda^2_{1,g}} \right)
%\frac{2}{\delta_{q}} 
- \frac{259}{60}- \frac{7f}{81} 
%-\ln \left(
%\frac{\Lambda^2_{1,g}}{Q^2} \right)
\right)  -
%\nonumber \\&-& 
\frac{8}{7}
 \frac{\Lambda^4_{2,g}}{Q^4} 
\left(
\ln \left(\frac{Q^2}{x^2\Lambda^2_{2,g}} \right)
%\frac{2}{\delta_{q}} 
- \frac{1817}{3360}- \frac{7f}{81} 
%-\ln \left(
%\frac{\Lambda^2_{2,g}}{Q^2} \right)
\right)
\Biggr\}
%\label{r7}
\nonumber
\eea

\begin{figure}[t]
%\rule{5cm}{0.2mm}\hfill\rule{5cm}{0.2mm}
%\vskip -2.5cm
\vskip -0.5cm
%\rule{5cm}{0.2mm}\hfill\rule{5cm}{0.2mm}
%\psfig{figure=filename.ps,height=1.5in}
%\psfig{figure=fi1.ps,height=1.5in}
\psfig{figure=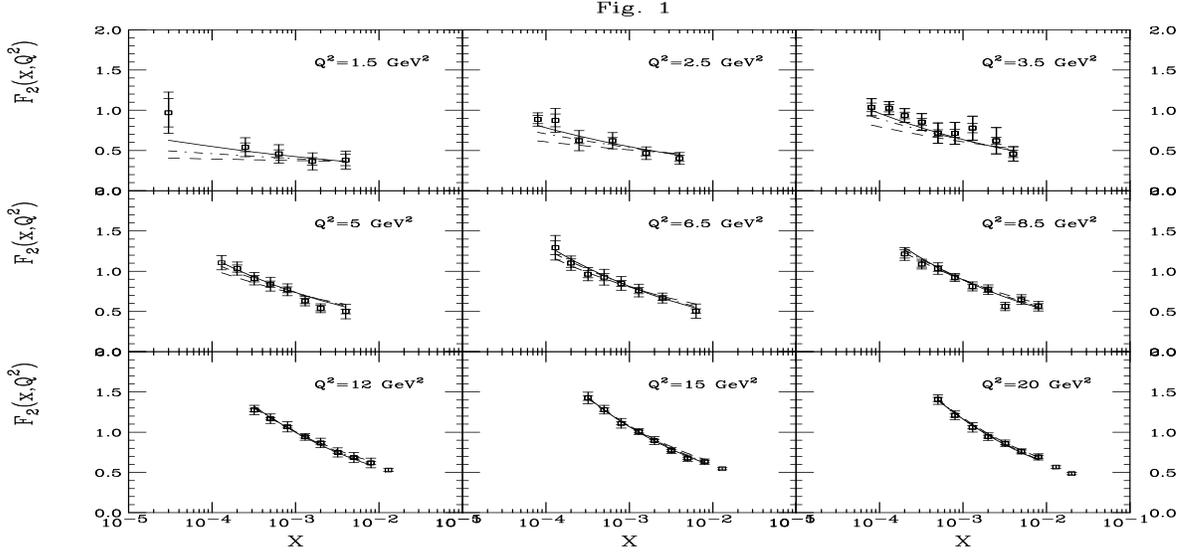,height=2.9in,width=6.2in}
%\psfig{figure=fi1h.ps,height=5.0in,width=5.in}
%\vskip -0.3cm
%
\caption{The structure function $F_2$ as a function of $x$ for different
$Q^2$ bins. The experimental points are from H1 \cite{H1}. 
The inner error 
bars are statistic while the outer bars represent statistic and systematic 
errors added in quadrature. The dashed and dot-dashed curves are obtained 
from fits (based on leading-twist formulae)
at LO and NLO respectively with fixed $Q^2_0=1$ GeV$^2$. The solid
line is from the fit at NLO giving $Q^2_0=0.55$ GeV$^2$.
% $\Lambda_{\ms}(f=4) = 250$ MeV is fixed.
}
%%%%\vskip -0.5cm
% \label{fig:radish}
\end{figure}

From Eqs.(\ref{r1})-(\ref{r7}) one can notice that the higher-twist
terms modify the flat condition Eq.(\ref{1}). They lead to a rise of
parton dostributions and, thus, $F_2$ structure function
at low value $Q^2_0$, when $x \to 0$. This is in agreement
with HERA data \cite{H1n}, as it is shown in next section.

%%\subsection{Fonts}

\section{Results of the fits}

With the help of the results presented in the previous section we have
analyzed $F_2$ HERA data at small $x$ from the H1 collaboration
(first articles in \cite{H1,H1n}).
%and ZEUS \cite{ZEUS} collaborations separately.
%Initially our solution of the DGLAP equations depends on five
%parameters, i.e. $Q_0^2$, $x_0$, $A_q$, $A_g$ and $\Lambda_{\ms}(n_f=4)$.
In order to keep the analysis as simple as possible
we have fixed the number of active flavors $f$=4 and
$\Lambda_{\ms}(n_f=4) = 250$ MeV, which
is a reasonable value extracted from the traditional (higher $x$)
experiments.
Moreover, we put $\Lambda_{1,a}=\Lambda_{2,a}$ in agreement with
\cite{DaWe}.
%and that has also been used by others \cite{Munich1}.
The initial scale of the 
%PD 
parton densities was also fixed
into the fits to $Q^2_0$ = 1 $GeV^2$, although later it was released
to study the sensitivity of the fit to the variation of this parameter.
The analyzed data region was restricted to $x<0.01$ to remain within the
kinematical range where our results are
accurate. 
%Finally, the number of active flavors was fixed to $f$=4. 

%\verb|\includegraphics{fi1h.ps }|

Fig. 1 shows $F_2$ calculated from the fit (based only
on leading-twist formulae)
with Q$^2$ $>$ 1 GeV$^2$
%given in table 1
in comparison with 1994 H1 data
(first article in \cite{H1}).
Only the lower $Q^2$ bins are shown.
One can observe that the NLO result (dot-dashed line)
lies closer to the data
than the LO curve (dashed line).
The lack of agreement between data and lines observed
at the lowest $x$ and $Q^2$ bins suggests
that the flat behavior should occur at $Q^2$ lower
than 1 GeV$^2$.
In order to study this point we have done the
analysis considering $Q_0^2$ as a free parameter.
Comparing the results of the fits (see \cite{Q2evo})
%in table 3 with those in table 2
one can notice
%a significant reduction in the value of
%$A_g$, $Q_0^2$ and the $\chi^2$. In Fig. 1
the better agreement with the experiment 
%of the NLO curve 
at fitted $Q^2_0=0.55$ GeV$^2$ (solid curve)
is apparent at the lowest kinematical bins.

\begin{figure}[t]
%\rule{5cm}{0.2mm}\hfill\rule{5cm}{0.2mm}
%\vskip -2.5cm
\vskip -0.5cm
%\rule{5cm}{0.2mm}\hfill\rule{5cm}{0.2mm}
%\psfig{figure=filename.ps,height=1.5in}
%\psfig{figure=fi1.ps,height=1.5in}
\psfig{figure=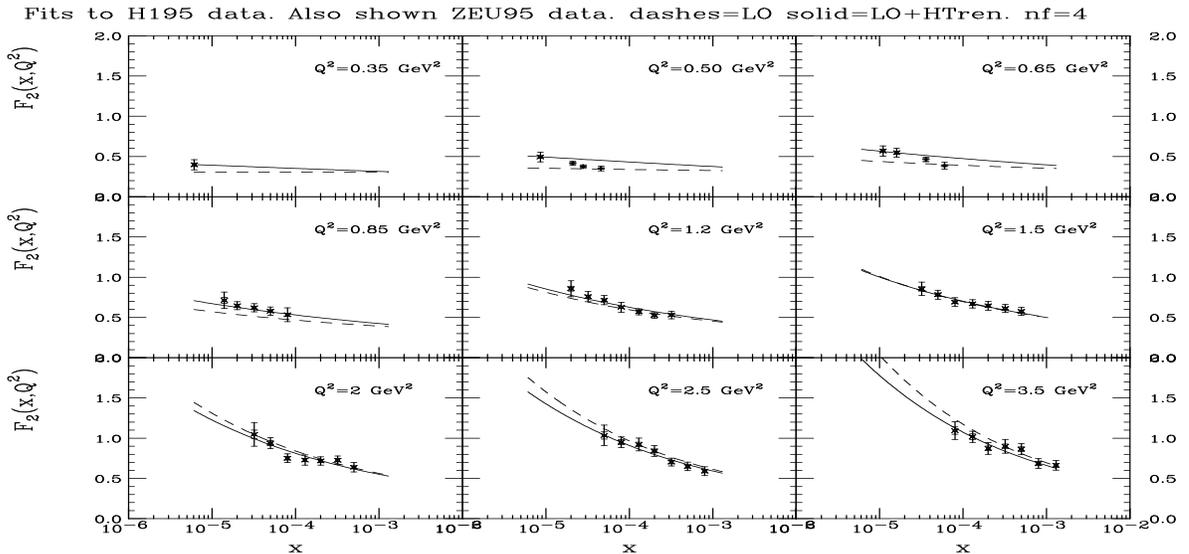,height=2.9in,width=6.2in}
%\psfig{figure=fi1h.ps,height=5.0in,width=5.in}
%\vskip -0.3cm
%
\caption{The structure function $F_2$ as a function of $x$ for different
$Q^2$ bins. The experimental points are from H1 and ZEUS \cite{H1n}. 
The inner error 
bars are statistic while the outer bars represent statistic and systimatic 
errors added in quadrature. The solid curves are obtained 
from fits at LO, when  contributions of 
higher-twist terms have been incorporated. 
The dashed curves are from the twist-two contributions alone.
%and NLO respectively with fixed $Q^2_0=1$ GeV$^2$. The solid
%line is from the fit at NLO giving $Q^2_0=0.55$ GeV$^2$.
% $\Lambda_{\ms}(f=4) = 250$ MeV is fixed.
}
%\vskip -0.5cm
% \label{fig:radish}
\end{figure}

Fig. 2 shows $F_2$ calculated from the fit at LO (based on leading-twist 
and higher-twist formulae)
%with Q$^2$ $>$ 1 GeV$^2$
%given in table 1
%%in comparison with 1995 H1 data
in comparison with 1995 H1 and ZEUS data \cite{H1n}.
%%(first article in \cite{H1n}).
%%The new low $Q^2$ ZEUS data (second article in \cite{H1n})
%%are also shown.
%Only the lower $Q^2$ bins are shown.
One can observe that these results 
%based on leading-twist and higher-twist terms 
%%dot-dashed 
(solid line)
lies closer to the data
than the twist-two results (dashed line).
We have done the
analysis considering $Q_0^2$ as a free parameter.
Comparing the results of the fits (see \cite{Q2evo3})
%in table 3 with those in table 2
one can notice
%a significant reduction in the value of
%$A_g$, $Q_0^2$ and the $\chi^2$. In Fig. 1
the better agreement with the experiment 
%of the NLO curve 
at fitted $Q^2_0=0.61$ GeV$^2$, 
%(solid curve)
%is apparent at the lowest kinematical bins.
%that is in agrrement with previous 
which is close to $Q_0^2$ in the
analysis of 1994 H1 data (see Fig. 1).

\section{Conclusions} 

We have shown that the results developed recently in \cite{Q2evo,Q2evo3}
%As we have shown, these results
 have quite simple form and reproduce many
properties of parton distributions at small $x$,
that have been known from global fits.

We found very good agreement between our approach based on QCD 
%at
%NLO approximation 
and HERA data, as it has been observed earlier with
other approaches (see the review \cite{CoDeRo}). 
%Thus, the nonperturbative
%contributions as shadowing effects,
%%\cite{Levin},
%higher twist effects
%%\cite{Bartels}
%and others seems to be quite small or seems to be canceled
%between them and/or with $ln(1/x)$ terms containing by higher orders of
%perturbative theory (see discussion also in \cite{Bartels}).
%In our opinion, this very good agreement between 
%%our 
%approaches based on perturbative QCD 
%%at NLO approximation 
%and HERA data may be explained also by the fact that
%at low $x$ values
%the real effective scale of coupling constant
%is like $Q^2/x^c$, where $1/2 \leq c \leq 1$
%(see \cite{scale}). To clear up the correct contributions of nonperturbative
%dynamics and higher orders containing large $ln(1/x)$ terms, it is
%necessary
%more precise data and further efforts in developing of theoretical
%approaches.
%
The (renormalon-type) higher-twist terms lead to the natural explanation of
the rise of $F_2$ structure function at low values of $Q^2$ and $x$.
The rise has been discovered in recent HERA experiments \cite{H1n}.

As next step of our investigations, we plan to study contributions of
higher-twist operators to relations between parton distributions and
deep inelastic structure functions, observed, for example, in 
\cite{KoPa,KOPAFL}.

  {\it Acknowledgments.}
One of the authors (A.V.K.)
%(A. V. K.)
 would like to express his sincerely thanks to the Organizing
 Committee and especially to Pavel Zarubin
%R. Fiore and A. Pappa 
for the kind invitation 
%and the financial support
at  such remarkable Conference.
%, and V.S. Fadin, L.L. Jenkovszky
%and L.N Lipatov 
%%and E.?. Martynov
%for fruitful discussions.
%%\\
%%One of the authors (A.V.K.) 
A.V.K. and G.P. 
were supported in part, respectively, by Alexander von Humboldt fellowship and
RFBR (98-02-16923) and
%and secon one (G.P.) was supported in part 
by Xunta de Galicia
(PXI20615PR) and CICYT (AEN99-0589-C02-02).

\end{document}